\documentclass{article}
\usepackage{amssymb,amsfonts,amsthm,bm,color}
\usepackage[round,numbers,sort&compress]{natbib}
\usepackage[ansinew]{inputenc}
\usepackage{mathtools}

\begin{document}

\title{The validity of quasi steady-state approximations in discrete stochastic simulations}

\author{Jae Kyoung Kim,$^1$ Kre\v simir Josi\' c,$^{2,3,\ast}$ and Matthew R.\ Bennett$^{4,5,\ast}$}

\maketitle

\noindent
{\footnotesize
{1}: Mathematical Biosciences Institute, The Ohio State University, Columbus, OH, 43210;\\
{2}: Department of Mathematics, University of Houston, Houston, TX, 77204;\\
{3}: Department of Biology and Biochemistry, University of Houston, Houston, TX, 77204;\\
{4}: Department of Biochemistry \& Cell Biology, Rice University, Houston, TX, 77005;\\
{5}: Institute of Biosciences and Bioengineering, Rice University, Houston, TX, 77005;\\
{*}: Correspondence: josic@math.uh.edu or matthew.bennett@rice.eduy \\}

\begin{abstract}
In biochemical networks, reactions often occur on disparate timescales and can be characterized as either ``fast'' or ``slow.'' The {\it quasi-steady state approximation} (QSSA) utilizes  timescale separation to project models of biochemical networks onto lower-dimensional slow manifolds. As a result, fast elementary reactions are not modeled explicitly, and their effect is captured by non-elementary reaction rate functions ({\it e.g.}\ Hill functions). The accuracy of the  QSSA applied to deterministic systems  depends on how well timescales are separated. Recently,  it has been proposed to use the non-elementary rate functions obtained via the deterministic QSSA to define propensity functions in stochastic simulations of biochemical networks. In this approach, termed the {\it stochastic QSSA}, fast reactions that are part of non-elementary reactions are not simulated, greatly reducing computation time. However, it is unclear when the stochastic QSSA provides an accurate approximation of the original stochastic simulation. We show that, unlike the deterministic QSSA, the validity of the stochastic QSSA does not follow from timescale separation alone, but also depends on the sensitivity of the non-elementary reaction rate functions to changes in the slow species. The stochastic QSSA becomes more accurate when this sensitivity is small. Different types of QSSAs result in non-elementary functions with different sensitivities, and the total QSSA 
results in less sensitive functions than the standard or the pre-factor QSSA. We prove that, as a result, the stochastic QSSA becomes more accurate when non-elementary reaction functions are obtained using the total QSSA. Our work provides a novel condition for the validity of the QSSA in stochastic simulations of biochemical reaction networks with disparate timescales. 
\end{abstract}

\markboth{Kim et al.}{QSSA in stochastic simulations}

\textbf{Note: This pre-print has been accepted for publication in Biophysical Journal. The final copyedited version of this paper will be available at www.biophyj.org }
\section*{Introduction}
In both prokaryotes and eukaryotes, the absolute number of a given reactant is generally small~\cite{Ghaemmaghami2003, Ishihama2008}, leading to high intrinsic noise in reactions. The Gillespie algorithm is widely used to simulate such reactions by generating sample trajectories from the chemical master equation (CME) \cite{Gillespie1977}. Because the Gillespie algorithm requires the simulation of every reaction, simulation times are dominated by the computation of fast reactions. For example, in an exact stochastic simulation most time is spent on simulating the fast binding and unbinding of transcription factors to their promoter sites, although these reactions are of less  interest than transcription which is slower. Thus, the Gillespie algorithm is frequently too inefficient to simulate biochemical networks with reactions spanning multiple timescales \cite{Gillespie2007, Cai2007}. 

\begin{figure}
\centerline{\includegraphics[width=3in]{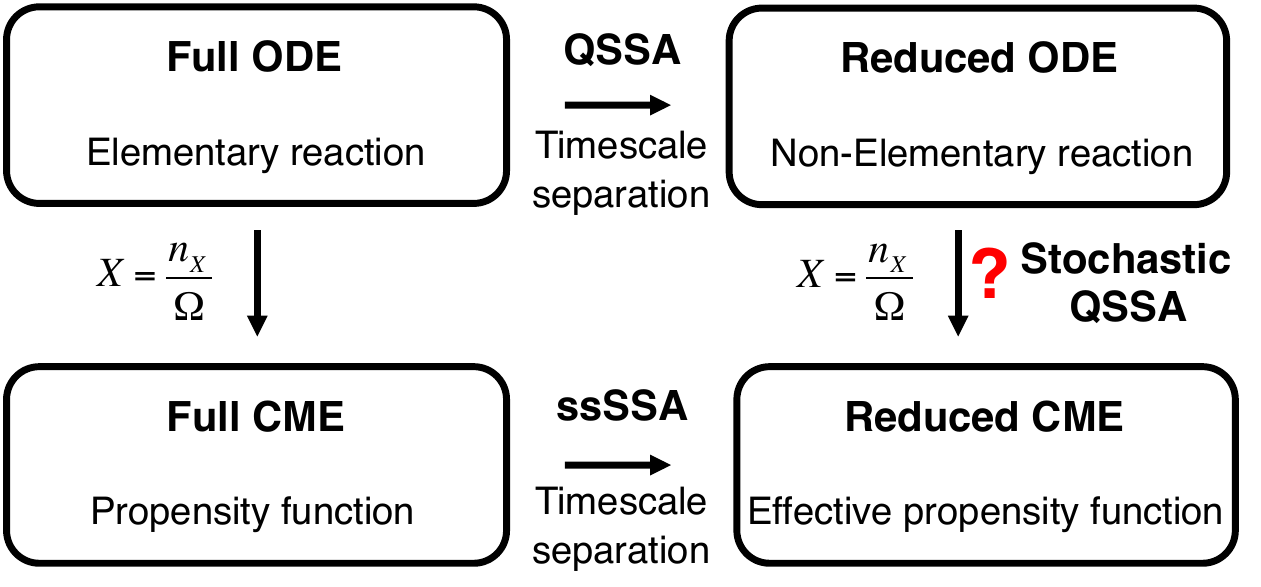}}
\caption{\label{fig:sum} The validity of the stochastic QSSA. Under timescale separation, the full ODE and full CME can be reduced using the deterministic QSSA and ssSSA, respectively. 
By changing the concentration ($X$) to the number of molecules ($n_X$) with the relationship $X=\frac{n_X}{\Omega}$ ($\Omega$: the volume of the system), the elementary rate functions in the full ODE can be converted to propensity functions. These are the same as the propensity functions of the full CME, which are derived from collision theory \cite{Gillespie1992}. However, the validity of propensity functions derived from non-elementary rate functions ({\it e.g.}\ Hill functions) of the reduced ODE is unclear. 
}
\end{figure}

Recently, the slow-scale stochastic simulation algorithm (ssSSA) was introduced to accelerate such simulations ~\cite{Gillespie2007, Cai2007} (Fig.\ \ref{fig:sum}). The main idea behind the ssSSA is to use the fact that fast species equilibrate quickly. Thus, we can replace fast species by their average values to derive effective propensity functions. These average values can be obtained by applying a quasi-steady-state approximation (QSSA) \cite{Rao2003, Barik2008, Macnamara2008} or quasi-equilibrium approximation \cite{Goutsias2005, Cao2005} to the CME. When using the ssSSA we only need to simulate slow reactions,  greatly increasing simulation speed with no significant loss of accuracy \cite{Gillespie2007, Cai2007, Rao2003, Barik2008, Macnamara2008, Goutsias2005, Cao2005}. However, the utility of the ssSSA is limited by the difficulty of calculating the average values of fast species, which requires knowledge of the joint probability distribution of the CME~\cite{Gillespie2007, Cai2007, Rao2003, Barik2008, Cao2005}. 

To estimate the averages value of the fast species, Rao {\it et al.}\ proposed using the fast species concentration at quasi-equilibrium in the deterministic system~\cite{Rao2003}. In such a {\it stochastic QSSA}, the deterministic QSSA is used to approximate the propensity functions obtained via the ssSSA (Fig.\ \ref{fig:sum}). Thus, non-elementary macroscopic rate functions ({\it e.g.}\ Hill functions) are used to derive the propensity functions in the same way as elementary rate functions ({\it i.e.}\ those obtained directly from mass action kinetics). Several numerical studies supported the validity of  the stochastic QSSA  in systems as diverse as Michaelis-Menten enzyme kinetics, bistable switches, and circadian clocks \cite{Gonze2002, Rao2003, Barik2008, Sanft2011}. These studies provided evidence that the stochastic QSSA is valid when timescale separation holds \cite{Thomas2011, Agarwal2012, Thomas2012}. Therefore, stochastic simulations of  biochemical networks are frequently  performed without converting the non-elementary reactions to their elementary forms \cite{Ouattara2010, Gonze2011a, Kim2013}. Moreover,  rates of the individual elementary reactions that are jointly modeled using Michaelis-Menten or Hill functions are rarely known, making the use of stochastic QSSA tempting. However, recent studies have demonstrated that, in contrast to the deterministic QSSA, timescale separation does not generally guarantee the accuracy of the stochastic QSSA~\cite{Thomas2011, Agarwal2012, Thomas2012}. The stochastic QSSA can often lead to large errors even when timescale separation holds. This raises the question: When is the stochastic QSSA valid?

Here, we investigate the conditions under which the stochastic QSSA is accurate (Fig.\ \ref{fig:sum}). We first  examine three of the most common reduction schemes: standard QSSA (sQSSA), total QSSA (tQSSA), and pre-factor QSSA (pQSSA). We find that the accuracy of the stochastic QSSA depends on which reduction scheme is used to derive its deterministic counterpart. Specifically, the stochastic tQSSA is more accurate than the stochastic sQSSA or pQSSA (we refer to each stochastic QSSA by the name of its deterministic counterpart, {\it i.e.} in the stochastic tQSSA, propensities are derived from the ODEs obtained via the deterministic tQSSA). All three methods relate the fast species concentration in quasi-equilibrium to the slow species concentration. For the tQSSA, this expression is  less sensitive to changes in the slow species than either its sQSSA or pQSSA counterparts. We find that for parameters that decrease sensitivity, the stochastic sQSSA and pQSSA also become more accurate. We explain these observations by proving that, as  sensitivity decreases, the approximate propensity functions used in the stochastic QSSA converge to the propensity functions obtained using ssSSA (Fig.\ \ref{fig:sum}).  Furthermore, we use a linear noise approximation (LNA) to show that the accuracy of the stochastic QSSA is determined by both separation of timescales {\it and} sensitivity. 

In sum, our results indicate that the stochastic QSSA is valid under more restrictive conditions than the deterministic QSSA. Importantly, we identify these conditions, and provide a theoretical foundation for reducing stochastic models of complex biochemical reaction networks with disparate timescales. 

\section*{Results}
\section*{The different types of deterministic QSSA}
\begin{figure*}
\begin{center}
\includegraphics[width=5in]{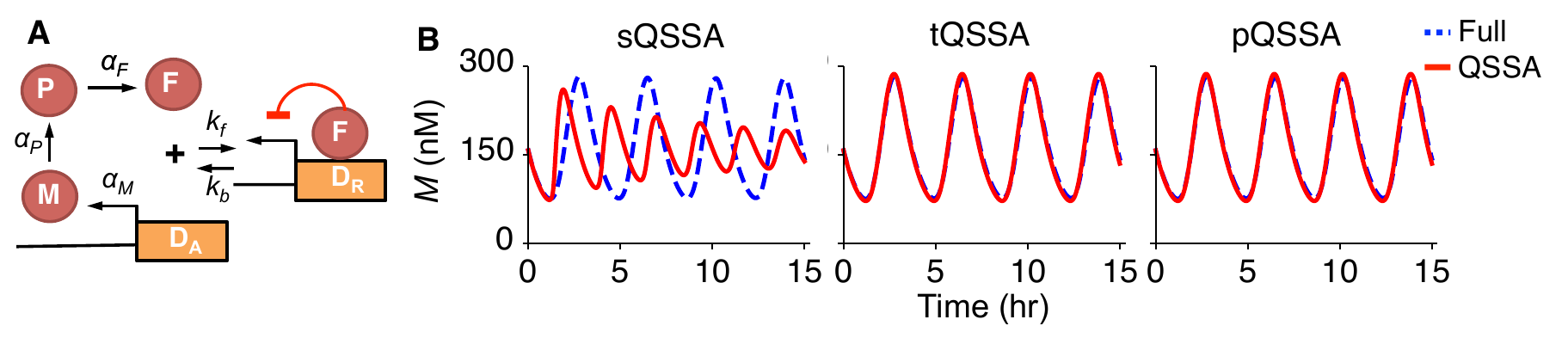}
\end{center}
\caption{\label{fig:det} Deterministic QSSA. (A) In the model of a genetic negative feedback loop, the repressor protein ($F$) represses its own transcription by binding to DNA promoter site ($D_A$). Reversible binding between $F$ and $D_A$ is much faster than other reactions. See Table S1 for parameters. (B) The sQSSA fails to correctly approximate the dynamics of the full system, but both the tQSSA and the pQSSA provide accurate approximations due to complete timescale separation between the variables.}
\end{figure*}

The term ``QSSA'' is used to describe a number of related dimensional reduction methods. We first review three common QSSA schemes using the example of a genetic negative feedback model \cite{Kim2012, Kim2014}. The full model, depicted in Fig.\ \ref{fig:det}A, can be described by the system of ODEs:
\begin{eqnarray}
\dot {M} &=& \alpha_M D_A - \beta_M M\label{eq:mdot}\\
\dot {P} &=&  \alpha_P M - \beta_P P\label{eq:rcdot}\\
\dot {F} &=&  \alpha_F P - \beta_F F -k_f F D_A + k_b D_R\label{eq:rfdot}\\
\dot {D_R} &=& k_f F D_A - k_b D_R - \beta_F D_R\label{eq:drdot}\\
\dot {D_A} &=& -k_f F D_A + k_b D_R + \beta_F D_R,\label{eq:dadot}\
\end{eqnarray}
where the transcription of mRNA ($M$) is proportional to the concentration of DNA promoter sites that are free of the repressor protein ($D_A$). The mRNA is translated into cytoplasmic protein ($P$). The free repressor protein ($F$) is produced at a rate proportional to the concentration of $P$. The free repressor can bind to a promoter site and change the DNA to its repressed state ($D_R$). All species, except for DNA, are subject to degradation ($\beta_i$), with the bound and free repressor degrading at the same rate. As can be seen in Eqs.\ \ref{eq:drdot} and \ref{eq:dadot}, total DNA concentration ($D_T=D_A + D_R$) is conserved. See Table S1 for the descriptions and values of parameters. 

\subsection*{Standard QSSA (sQSSA)}
Binding ($k_f$) and unbinding ($k_b$) between $F$ and $D_A$ are much faster than the remaining reactions (Fig.\ \ref{fig:det}A and Table S1). Thus, Eqs.\ \ref{eq:drdot} and \ref{eq:dadot} equilibrate faster than Eqs.\ \ref{eq:mdot}-\ref{eq:rfdot}, which leads QSS equations for the fast species ($\dot {D_R}=0$ and $\dot {D_A}=0$). By solving these QSS equations, we obtain the equilibrium values of fast species ($D_R$ and $D_A$) in terms of slow species ($F$):
\begin{eqnarray}
D_R (F) &=& \frac {D_T F}{F + K_d}\label{eq:drrf}\\
D_A (F) &=& \frac {D_T K_d}{F + K_d},\label{eq:darf}\
\end{eqnarray}
where $K_d=(k_b + \beta _F) /k_f$. These QSS solutions can be used to close the remaining equations (Eqs.\ \ref{eq:mdot}-\ref{eq:rfdot}) giving the reduced system:
\begin{eqnarray}
\dot {M} &=& \alpha_M \frac {D_T K_d}{F + K_d} - \beta _M M\label{eq:mdots}\\
\dot {P} &=&  \alpha_P M - \beta_P P\label{eq:rcdots}\\
\dot {F} &=&  \alpha_F P - \beta_F \left(F + \frac {D_T F}{F + K_d}\right).\label{eq:rfdots}\
\end{eqnarray}
This approach is known as the classical or standard QSSA (sQSSA) \cite{Menten1913, Segel1989, Schnell2000}. Previous studies have shown that the sQSSA leads to reductions that correctly predict steady-states, but may not correctly describe the dynamics~\cite{Kepler2001,Bennett2007}.  Indeed, whereas the original system (Eqs.\ \ref{eq:mdot}-\ref{eq:dadot}) relaxes to a limit cycle, the reduced system (Eqs.\ \ref{eq:mdots}-\ref{eq:rfdots}) exhibits damped oscillations  (Fig.\ \ref{fig:det}B).

\subsection*{Total QSSA (tQSSA)}
The inaccuracy of the sQSSA results from treating $F$ as a slow variable, even though it is affected by both slow (production and degradation) and fast (binding and unbinding to DNA) reactions \cite{Bennett2007}. This problem can be solved by introducing the total amount of repressor, $R \equiv F + D_R$,  instead of $F$. As a result, $\dot R$ only depends on slow reactions:
\begin{eqnarray}
\dot M &=& \alpha_M D_A - \beta _M M\label{eq:mdottf}\\
\dot P &=&  \alpha_P M - \beta _P P\label{eq:rcdottf}\\
\dot R &=& \alpha_F P - \beta _F R\label{eq:rdottf}\\
\dot D_R &=& k_f (R-D_R) D_A - k_b D_R - \beta _F D_R\label{eq:drdottf}\\
\dot D_A &=& -k_f (R-D_R) D_A + k_b D_R + \beta _F D_R.\label{eq:dadottf}\
\end{eqnarray}
By solving the QSS equations for the fast species ($\dot {D_R}=0$ and $\dot {D_A}=0$), we obtain the equilibrium values of $D_R$ and $D_A$ in terms of $R$: 
 \begin{eqnarray}
D_R (R) &=& \frac{1}{2}\left(D_T + R + K_d -\sqrt{(D_T - R - K_d)^2 + 4D_T K_d}\right)\label{eq:drr}\\
D_A (R) &=& \frac{1}{2}\left(D_T - R - K_d+ \sqrt{(D_T - R - K_d)^2 + 4D_T K_d}\right).\label{eq:dar}
\end{eqnarray}
Substituting these QSS solutions to close the remaining equations (Eqs.\ \ref{eq:mdottf}-\ref{eq:rdottf}), we arrive at the reduced system 
\begin{eqnarray}
\dot M &=& \alpha_M D_A (R) - \beta_M M\label{eq:mdott}\\
\dot P &=&  \alpha_P M - \beta_P P\label{eq:rcdott}\\
\dot R&=& \alpha_F P - \beta_F R.\label{eq:rdott}\
\end{eqnarray}
This approach is known as the total QSSA (tQSSA) \cite{Tzafriri2003, Ciliberto2007, Kumar2011}. Due to the 
complete timescale separation between variables, the tQSSA leads to a reduced system (Eqs.\ \ref{eq:mdott}-\ref{eq:rdott}) that correctly captures the dynamics of the full system (Fig.\ \ref{fig:det}B). However, unlike the recognizably Michaelis-Menten-like form of sQSS solutions (Eqs.\ \ref{eq:drrf} and \ref{eq:darf}), the corresponding tQSS solutions (Eqs.\ \ref{eq:drr} and \ref{eq:dar}) are unfamiliar and unintuitive.

\subsection*{Pre-factor QSSA (pQSSA)}
The reduced system obtained with the tQSSA can be transformed into a more intuitive form. Expressing
 Eqs.\ \ref{eq:mdott}-\ref{eq:rdott} using the original free protein variable, $F$, and using  $\dot {R}=\frac{\partial R}{\partial F} \dot {F}$, we obtain:
\begin{eqnarray}
\dot {M} &=& \alpha_M \frac {D_T K_d}{F + K_d} - \beta _M M\label{eq:mdotp}\\
\dot {P} &=&  \alpha_P M - \beta _P P\label{eq:rcdotp}\\
p(F) \dot {F} &=& \alpha_F P - \beta _F \left(F + \frac {D_T F}{F + K_d}\right),\label{eq:rfdotp}\
\end{eqnarray}
where
\begin{equation}\label{eq:pref}
p(F) \equiv \frac{\partial R}{\partial F} = \frac{\partial F}{\partial F}+\frac{\partial D_R}{\partial F}=1+  \frac {D_T K_d}{(F + K_d)^2}.
\end{equation}
This approach is known as the pre-factor QSSA (pQSSA)~\cite{Kepler2001, Bennett2007}. We note two important things about Eqs.\ \ref{eq:mdotp}-\ref{eq:rfdotp}. First, the system is identical to that obtained using the sQSSA (Eqs.\ \ref{eq:mdots}-\ref{eq:rfdots}), except for the prefactor $p(F)$. Therefore, the two reductions have the same fixed points, but their dynamics are different. The pre-factor is always greater than one, and corrects the inaccuracy in the dynamics that are introduced in the sQSSA (Fig.\ \ref{fig:det}B). Second, because the pQSSA and tQSSA lead to equivalent systems (Eqs.\ \ref{eq:mdotp}-\ref{eq:rfdotp} and Eqs.\ \ref{eq:mdott}-\ref{eq:rdott}), the resulting dynamics are identical, up to a change of variables. In sum, due to complete timescale separation between variables, reduced ODE models obtained using the tQSSA or the pQSSA approximate the dynamics of the original system more accurately than the sQSSA. 

\section*{Stochastic QSSA}
We have derived the reduced system of a genetic negative feedback model (Eqs. \ref{eq:mdot}-\ref{eq:dadot}) using three types of the QSSA. These different reductions result in different propensity functions in the stochastic QSSA. We now investigate how the accuracy of the stochastic QSSA depends on the choice of the reduction. 

For discrete stochastic simulations, we need to convert the concentration of a reactant to the absolute number of molecules (Fig.\ \ref{fig:sum}). For instance, the concentration of mRNA, $M$, and the number of mRNA molecules, $n_M$, are related by $M=n_M / \Omega$, where $\Omega$ represents the volume of the system. In this study, we choose $\Omega=1$ for simplicity, so that the numerical values of the concentration and the number of molecules are equal. Using this type of relation, we obtain the propensity functions of the reactions from the corresponding macroscopic rate functions of the full and three reduced ODE models (Tables S2-5). The results of stochastic simulations with these propensity functions are shown in Fig.\ \ref{fig:sto}A. Similar to the deterministic simulations (Fig.\ \ref{fig:det}B), the simulations using the stochastic sQSSA exhibit faster oscillations than the full system, and simulations using the stochastic tQSSA correctly predict the dynamics of the full system (Fig.\ \ref{fig:sto}A). The deterministic reductions obtained using the tQSSA and the pQSSA are equivalent (Fig.\ {\ref{fig:det}B).  This suggests that their stochastic counterparts will also behave similarly. However, this is not the case:  Simulations using the stochastic pQSSA do not provide an accurate approximation of the full system (Fig.\ \ref{fig:sto}A). In particular, the fraction of active DNA, $n_{D_A}/n_{D_T}$, which determines the transcription rate of mRNA, exhibits large jumps when using the stochastic sQSSA and pQSSA, in contrast to the stochastic tQSSA (Fig.\ \ref{fig:sto}A).

This surprising behavior of $n_{D_A}/n_{D_T}$ when using the stochastic sQSSA and pQSSA is a result of the sensitive dependence of this ratio on the number of free repressor, $n_F$:
\begin{equation}
\frac {n_{D_A }} {n_{D_T}} = \frac {K_d \Omega}{n_F + K_d \Omega} \thickapprox 
\left\{ \begin{array}{ll}
 1 & n_F =0 \\
 0.2 & n_F =1\\
 0.11 & n_F  =2\\
 \vdots & \vdots
 \end{array}, \right.\label{eq:dss}\
\end{equation}
which is derived from the non-elementary form of the sQSS solution (Eq.\ \ref{eq:darf}). Only a few molecules of transcription factor are needed to strongly repress transcription. Therefore, when the QSS solution (Eq.\ \ref{eq:darf}) is used to derive $n_{D_A}/n_{D_T}$ (Eq.\ \ref{eq:dss}) in the case of the stochastic sQSSA or pQSSA, the stochastic simulations become extremely sensitive to fluctuations in $n_F$ when $n_F$ is small. This is the cause of the large jumps seen in Fig.\ \ref{fig:sto}A and the disagreement between the dynamics of the reduced and the original system.  The stochastic pQSSA leads to additional errors because the pre-factor  defined by Eq.\ \ref{eq:pref} is also sensitive to fluctuations in $n_F$ (Fig.\ \ref{fig:sto}A). 

However, in the stochastic tQSSA, the ratio $n_{D_A}/n_{D_T}$, which is derived from the tQSS solution (Eq.\ \ref{eq:dar}), is less sensitive to changes in the total amount of repressor, $n_R$:  
 \begin{eqnarray}
\frac {n_{D_A}} {n_{D_T}} &=& \frac{1}{2 n_{D_T}} \left(n_{D_T}- n_R  - K_d \Omega +\sqrt{\left(n_{D_T} - n_R - K_d \Omega\right)^2 + 4 K_d \Omega n_{D_T}}\right)\nonumber\\
&\thickapprox&\left\{ \begin{array}{ll}
 1 & n_R =0 \\
 0.993939 & n_R=1\\
 0.98789 & n_R =2\\
 \vdots & \vdots
 \end{array} \right.\label{eq:dst}\
 \end{eqnarray}
As a result, the ratio $n_{D_A}/n_{D_T}$ does not exhibit large jumps, and the dynamics of the original system are approximated accurately when using the stochastic tQSSA (Fig.\ \ref{fig:sto}A).

The sensitivity of the ratio $n_{D_A} /n_{D_T}$  to changes in $n_F$ depends on system parameters. We expect that when this sensitivity is small, the stochastic sQSSA or pQSSA become more accurate. One way to reduce such sensitivity is to increase  $K_d$ in Eq.\ \ref{eq:dss}. As $K_d$ increases, the deterministic system ceases to oscillate and asymptotically approaches a fixed point, so that  we can measure the coefficient of variation (CV) of $n_M$ at equilibrium to describe the variability in the system. As shown in Fig.\ \ref{fig:sto}B, as $K_d$ increases and the sensitivity of Eq.\ \ref{eq:dss} decreases, the stochastic sQSSA and pQSSA become more accurate. Furthermore, the stochastic tQSSA is accurate at all values of $K_d$ due to the low sensitivity of Eq.\ \ref{eq:dst}.

\begin{figure*}
\begin{center}
\includegraphics[width=5in]{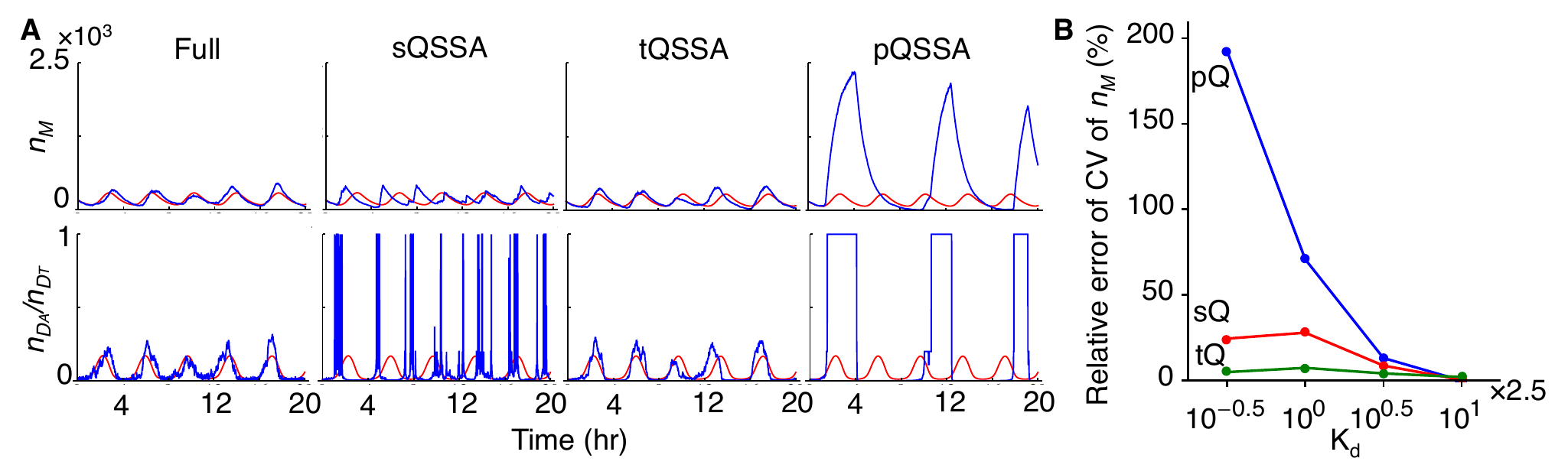}
\end{center}
\caption{\label{fig:sto} Stochastic QSSA. (A) Whereas the deterministic pQSSA and tQSSA are equivalent (Fig.\ \ref{fig:det}B), simulation results of the stochastic pQSSA and tQSSA do not agree. In particular, the amount  of active DNA, described by Eqs.\ \ref{eq:dss} and \ref{eq:dst}, exhibits large jumps when using the sQSSA and the pQSSA, but not the tQSSA. Results of the deterministic simulation of the full system are shown in red, while results of the stochastic simulations are shown in blue. (B) As $K_d$ increases, the sensitivity of the QSS solution (Eq.\ \ref{eq:dss}) decreases, which results in more accurate simulations of the stochastic QSSA. The coefficient of variation (CV) of mRNA $(n_M)$ at its steady state is estimated with 25,000 independent simulations for each system. Each simulation is run until 20 hr of reaction time to ensure the system in the stationary state.}
\end{figure*}

\section*{The accuracy of the stochastic QSSA depends on the sensitivity of the QSS solution}
We next provide a more complete analysis of the relationship between the sensitivity of the QSS solution and the accuracy of the stochastic QSSA. In our model, the reversible binding between free repressor protein and DNA,
\begin{equation}\label{eq:bu}
F+D_{A} {\longleftrightarrow} D_R,
\end{equation}
is much faster than other reactions. The amount of active DNA is governed by this fast reaction and determines the dynamics of the slow process, specifically the transcription of mRNA with propensity function $\alpha_{M}n_{D_{A}}$. Previous studies have shown that, assuming timescale separation, this propensity function can be approximated by an effective propensity function, $\alpha_{M}\left< n_{D_{A}}\right>$~\cite{Gillespie2007, Cai2007, Cao2005}. This approach is known as the ssSSA (Fig.\ \ref{fig:sum}). 
Here, the average, $\left< \cdot \right>,$ is defined by
\begin{equation}\label{eq:avedef}
\left<x\right>\equiv\sum_{x=0}^\infty xP(x|\mathbb{S}),
\end{equation}
where $P(x|\mathbb{S})$ is the stationary probability distribution of $x$ given a fixed state, $\mathbb{S},$ of slow species. That is, we compute the average of the fast species in quasi-equlibrium. Hence, $\left< n_{D_{A}}\right>$ is the mean of the steady-state distribution of active DNA evolving only through fast reactions, with slow species ``frozen'' in time. The main idea behind the ssSSA is that $n_{D_{A}}$ quickly relaxes to $\left< n_{D_{A}}\right>$, so that over slow timescales $n_{D_{A}}$ can be replaced with  $\left< n_{D_{A}}\right>$ (Fig. \ref{fig:sum}) \cite{Gillespie2007, Cai2007, Cao2005}. However, $\left< n_{D_{A}}\right>$ is usually unknown, so the stochastic QSSA approximates $\left< n_{D_{A}}\right>$ with a QSS solution. One can estimate the error in using either the sQSS solution $n_{D_{A}}(n_F)$ (Eq.\ \ref{eq:dss}) or the tQSS solution $n_{D_{A}}(n_R)$ (Eq.\ \ref{eq:dst}) to approximate $\left< n_{D_{A}}\right>$ by equating moments~\cite{Sanft2011} (see supplementary information for details). For the stochastic sQSSA and pQSSA this leads  to 
\begin{equation}\label{eq:mset}
\left<n_{D_{A}}\right> \approx n_{D_{A}}(\left<n_F\right>)+\frac{Var(n_{D_{A}})}{n_{D_{A}}(\left<n_F\right>)}\frac{dn_{D_{A}}(\left<n_F\right>)}{d\left<n_F\right>},
\end{equation}
and for the stochastic tQSSA we arrive at
\begin{equation}\label{eq:mtet}
\left<n_{D_{A}}\right> \approx n_{D_{A}}(n_R)+\frac{Var(n_{D_{A}})}{n_{D_{A}}(n_R)} \frac{dn_{D_{A}}(n_R)}{dn_R}.
\end{equation}
Here, $n_{D_{A}}(\left<n_F\right>)$ in Eq.\ \ref{eq:mset} agrees with the expression for $n_{D_A}(n_F)$ derived from the sQSS solution (Eq.\ \ref{eq:dss}) because $\left< n_F\right>$ approximates $n_F$ under slow timescale. The errors of both the sQSS and tQSS solutions above depend on the Fano factor of the fast species, $\frac{Var(n_{D_{A}})}{n_{D_{A}}}$, because the QSS solutions agree with $\left<n_{D_{A}}\right> $ under the moment closure assumption (see supplementary information for details). That is, the error in the stochastic QSSA arises mainly from ignoring the variance of fast variables, which will vanish along with random fluctuations in the limit of large system size. Interestingly, the magnitude of the error  depends on the sensitivity of the QSS solution. In particular, $\frac{dn_{D_{A}}}{dn_{R}}$, the sensitivity of the tQSS solution (Eq.\ \ref{eq:dst}), is small  because $| \frac{dn_{D_{A}}}{dn_R} | =| \frac{dn_{D_{R}}}{dn_R} |<1$ regardless of parameter choice. This explains the accuracy of the stochastic tQSSA (Fig. \ref{fig:sto}). However, the sensitivity of the sQSS solution ($\frac{dn_{D_{A}}(\left<n_F\right>)}{d\left<n_F\right>}$) can be large (Eq.\ \ref{eq:dss}). Eq.\ \ref{eq:mset} implies that the accuracy of the stochastic sQSSA and pQSSA deteriorates as $\frac{dn_{D_{A}}(\left<n_F\right>)}{d\left<n_F\right>}$ increases, which explains our previous simulation results (Fig.\ \ref{fig:sto}). From Eqs.\ \ref{eq:mset} and \ref{eq:mtet}, we can also compare the errors of the two approximations obtained with the sQSSA and the tQSSA:
\begin{eqnarray} 
\frac{\left<n_{D_{A}}\right>-n_{D_{A}}(\left<n_F\right>)}{\left<n_{D_{A}}\right>-n_{D_{A}}(n_R)} \approx \frac{dn_{D_{A}}(\left<n_F\right>)}{d\left<n_F\right>}\big/ \frac{dn_{D_A}(n_R)}{dn_R} =\frac{dn_R}{d\left<n_F\right>} =1+\frac{d\left<n_{D_{r}}\right>}{d\left<n_F\right>}>1. \label{eq:mra}
\end{eqnarray}
This inequality follows from the observation that $\left<n_{D_R}\right>$ increases monotonically with $\left<n_F\right>$. Eq.\ 31 indicates that the tQSSA provides a better estimate of $\left<n_{D_{A}}\right>$ than the sQSSA or the pQSSA. More generally, the tQSS solution has lower sensitivity than sQSS or pQSS solutions if the components of the total variable used in the tQSSA have a positive, monotonic relationship with the variable used in the sQSSA or the pQSSA. That is, let us assume that $T=T_1 + T_2 + ... +T_n$ is the total variable used in the tQSSA (e.g.\ $R$) and $T_1$ is the slow variable used for the sQSSA and the pQSSA (e.g.\ $F$). If $\frac{dT_i}{dT_1}>0$ for all $i=2,...,n$, then the tQSS solution always has lower sensitivity than the sQSS solution or the pQSS solution. Widely used QSS solutions, such as Hill-functions, satisfy this condition.

In summary, the non-elementary form of the QSS solutions derived using the sQSSA and the tQSSA provide estimates of the first moment of the fast species under a moment closure assumption, but with different choices of coordinates (Fig.\ \ref{fig:sum}). The error introduced by truncating higher moments depends on the sensitivity of the QSS solutions in both cases. These results are generalized to any system in which reversible binding reactions are faster than other reactions. The proof of the following theorem can be found in the supplementary information.\\

\textbf{Theorem}. Assume that a  biochemical reaction network includes a reversible binding reaction with a dissociation constant $K_d =k_b/k_f$, 
\begin{equation}
S+F {\longleftrightarrow} C.
\end{equation}
that is faster than the other reactions in the system.
Let $T\equiv S+C$ and $U\equiv F+C$. If $Var(n_C)\ll n_T n_U$, then $\left<n_C\right>$ and $\left<n_F\right>$ satisfy: 
\begin{eqnarray}
\left<n_C\right> &\approx& n_C(n_T)+\frac{Var(n_C)}{n_F(n_T)} \frac{dn_C(n_T)}{dn_T}\label{eq:tes}\\  
\left<n_F\right> &\approx& n_F(n_T)+\frac{Var(n_F)}{n_F(n_T)} \frac{dn_F(n_T)}{dn_T},\
\end{eqnarray}
where $n_C(n_T)$ is the solution of the tQSS equation, $n_C^2-(n_U+n_T+K_d \Omega)n_C+n_Un_T=0$, and  $n_F(n_T)=n_U-n_C(n_T)$. 
Similarly,
\begin{eqnarray}
\left<n_C\right> &\approx& n_C(\left<n_S\right>)+\frac{Var(n_C)}{n_F(\left<n_S\right>)}\frac{dn_C(\left<n_S\right>)}{d\left<n_S\right>}\label{eq:ses}\\
\left<n_F\right> &\approx& n_F(\left<n_S\right>)+\frac{Var(n_F)}{n_F(\left<n_S\right>)}\frac{dn_F(\left<n_S\right>)}{d\left<n_S\right>},\label{eq:sest}\
\end{eqnarray}
where $n_C(\left<n_S\right>)$  is the solution of the sQSS equation, $(\left<n_S\right>+K_d\Omega)n_F + n_U\left<n_S\right>=0$, and $n_F(\left<n_S\right>)=n_U-n_C(\left<n_S\right>)$.


\section*{Michaelis-Menten enzyme kinetics}
We first apply our theorem to Michaelis-Menten enzyme kinetics \cite{Menten1913,Briggs1925} under the assumption that the product of the  reaction can revert back to substrate. This example was recently used to explore the accuracy of the stochastic sQSSA \cite{Agarwal2012}. The deterministic model is described by:
\begin{eqnarray}
\dot S&=&-k_f SE + k_b C + k_s P\label{eq:ezfs}\\
\dot C&=&k_f SE - k_b C - k_p C\label{eq:ezf}\\
\dot P&=& k_p C - k_s P,\label{eq:ezfp}\
\end{eqnarray}
where the total enzyme concentration, $E_T \equiv C + E$, is constant. In this system, the free enzyme ($E$) reversibly binds substrate ($S$) to form the complex ($C$). The complex irreversibly dissociates into product ($P$) and free enzyme. The products can be converted back to substrate, and hence the substrate concentration is not equal to zero in steady state. We assume that binding ($k_f$) and unbinding ($k_b$) between $S$ and $E$ are much faster than the other reactions (see Table S6 for the details of parameters). Then, using conservation, $S_T \equiv S+C+P$ and solving the QSS equation ($\dot C=0$), we obtain the sQSSA system, 
\begin{equation}\label{eq:ezs}
\dot{S} =-k_p C(S) + k_s (S_T -S-C(S)),
\end{equation}
where $C(S)=\frac {E_T S}{K_m + S}$ and $K_m=(k_b+k_p)/k_f$. Next, if we define $T\equiv S+C$, we obtain the tQSSA system, 
\begin{equation}\label{eq:ezt}
\dot T=-k_p C(T) + k_s (S_T-T),
\end{equation}
where $C(T)=\frac{E_T +K_m +T -\sqrt{(E_T + K_m + T)^2 -4E_T T}}{2}$. In the stochastic QSSA, by chaining the concentration to the number of molecules in these QSS solutions (Eqs \ref{eq:ezs} and \ref{eq:ezt}), we approximate the average of fast species at quasi-equilibrium ($\left<n_C\right>$). Then, we can derive the relative errors of these approximations according to Eqs.\ \ref{eq:tes} and \ref{eq:ses}: 
\begin{eqnarray}
\displaystyle \frac{\left<n_C\right>-n_C(n_T)}{\left<n_C\right>} &\approx& \frac{1}{\left<n_C\right>}\frac{Var(n_C)}{n_{E}(n_T)} \frac{dn_C(n_T)}{dn_T},\label{eq:ezet}\\
\displaystyle \frac{\left<n_C\right>-n_C(\left<n_S\right>)}{\left<n_C\right>} &\approx& \frac{1}{\left<n_C\right>}\frac{Var(n_C)}{n_{E}(\left<n_S\right>)}\frac{dn_C(\left<n_S\right>)}{d\left<n_S\right>}\label{eq:ezes}.
\end{eqnarray} 

Similar to Eq.\ \ref{eq:mra}, $\frac{dn_C/dn_S}{dn_C/dn_T}>1$ regardless of parameter choice. For illustration we select two sets of parameters: for the first $\frac{dn_C/dn_S}{dn_C/dn_T} \approx 1$ (Fig.\ \ref{fig:enz}A), and for the second $\frac{dn_C/dn_S}{dn_C/dn_T} > 1$ (Fig.\ \ref{fig:enz}B). As expected from Eqs.\ \ref{eq:ezet}-\ref{eq:ezes}, with the first choice of parameters, tQSS and sQSS solutions give comparable results  in estimating $\left<n_C\right> $ (Fig.\ \ref{fig:enz}C). With the second parameter set, the sQSS solution leads to much larger errors than the tQSS solution (Fig.\ \ref{fig:enz}D). Furthermore, Eq.\ \ref{eq:ezet} and \ref{eq:ezes} predicts that the error ratio depends on the ratio of sensitivities of the sQSS and tQSS solutions ($\frac{dn_C/dn_S}{dn_C/dn_T}$). This prediction is supported by our simulations (Fig.\ \ref{fig:enz}E and F). Along with successful estimation of  $\left<n_C\right> $ when parameters are chosen so that $\frac{dn_C/dn_S}{dn_C/dn_T} \approx 1$, the stochastic simulations of slow variables using both the sQSSA (Eq.\ \ref{eq:ezs}) and the tQSSA (Eq.\ \ref{eq:ezt}) become accurate (Fig.\ \ref{fig:enz}G). However, for the parameters such that $\frac{dn_C/dn_S}{dn_C/dn_T} \gg 1$, the stochastic sQSSA results in much larger error than the stochastic tQSSA (Fig.\ \ref{fig:enz}H).

\begin{figure}
\centerline{\includegraphics[width=3.3in]{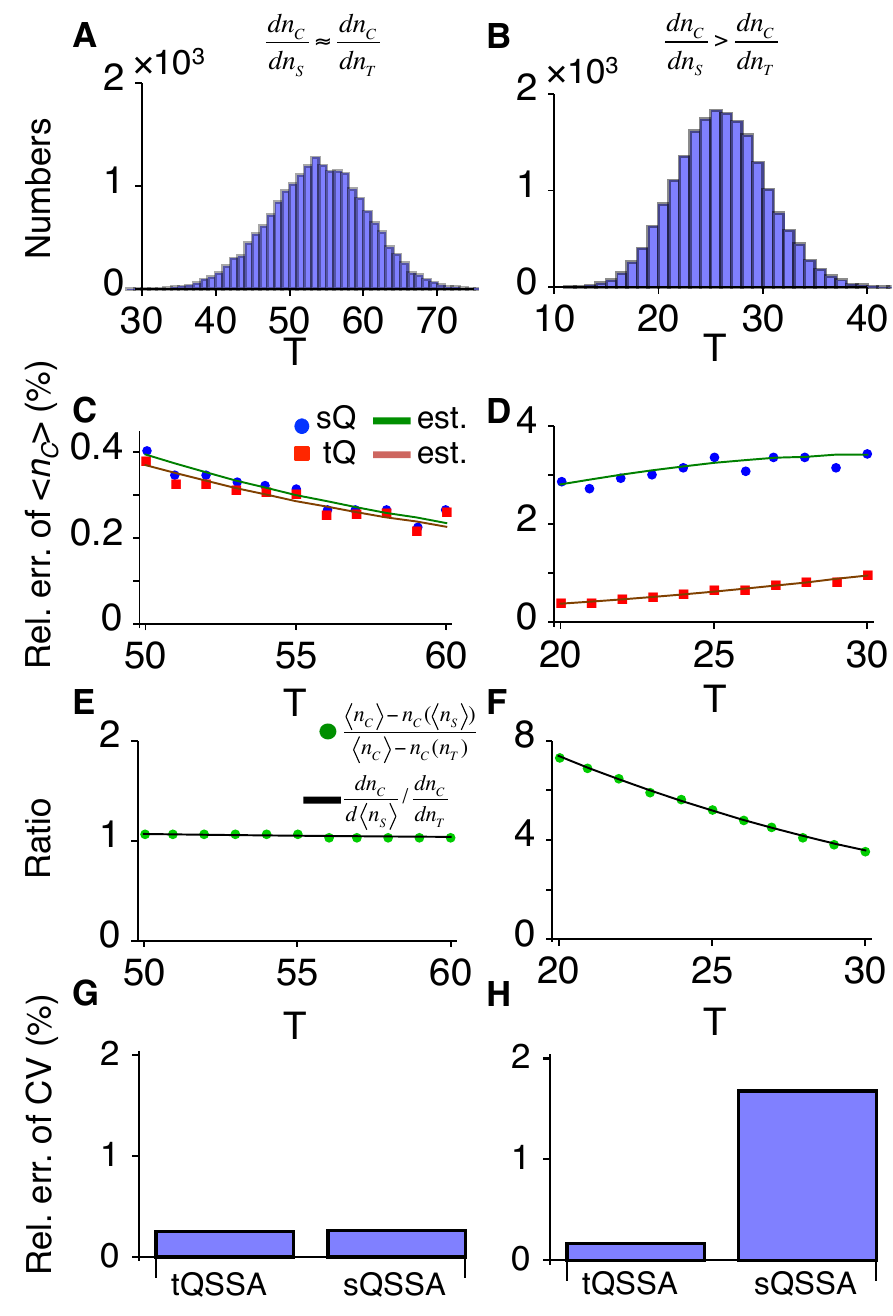}}
\caption{\label{fig:enz}  The stochastic QSSA of enzyme kinetics. (A, B) The distribution of $T$ of $20,000$ independent stochastic simulations of the full model (Eqs.\ \ref{eq:ezfs} - \ref{eq:ezfp}) for parameter sets having similar sensitivities (A) and different sensitivities (B) of sQSS and tQSS solutions (see Table S6 for parameters). Each simulation is run for 15000s of reaction time to ensure the system is in equilibrium. (C, D) The errors of QSS solutions used to  approximate $\left<n_C\right>$ for a given $T$. Blue circles and red squares indicate relative errors of the tQSS solution (left side of Eq. \ref{eq:ezet}) and the sQSS solution (left side of Eq. \ref{eq:ezes}), respectively. Orange and green line indicate our estimates  of the relative errors (right sides of Eqs. \ref{eq:ezet} and \ref{eq:ezes}). (E, F) The ratio between the errors with sQSS and tQSS solutions in Fig. \ref{fig:enz}C and D matches the ratio between the sensitivities of the sQSS and tQSS solutions. (G, H) Relative errors of the CV of the slow species when the stochastic sQSSA (Eq. \ref{eq:ezs}) and tQSSA (Eq. \ref{eq:ezt}) are used.}
\end{figure}

\section*{Genetic negative feedback loop with protein dimerization}
Next, we consider a more complex system that includes multiple fast reversible binding reactions. We adopt a model of the $\lambda$ repressor protein cI of phage $\lambda$ in \textit{E. coli} \cite{Bundschuh2003a,Bundschuh2003}, in which a dimeric protein represses its own transcription. The slow reactions in the model consist of transcription, translation, and degradation:
\begin{eqnarray}
D_A &\xrightarrow{\alpha_M}& D_A + M\\
M &\xrightarrow{\alpha_P}& M + P\\
M &\xrightarrow{\beta_M}& \phi\\
P &\xrightarrow{\beta_P}& \phi,\
\end{eqnarray}
where $D_A$ is free DNA, $M$ is mRNA, and $P$ is monomeric protein. See Table S7 for the details of parameters. The fast reactions of the model are the dimerization of monomer and the binding of the dimer to the DNA, 

\begin{eqnarray}
P + P &\xleftrightarrow[\text {$k_{b1}$}]{\text{$k_{f1}$}}& P_2\\
P_A + D_A &\xleftrightarrow[\text {$k_{b2}$}]{\text{$k_{f2}$}}& D_R,\
\end{eqnarray}
where $P_2$ is dimeric protein and $D_R$ is DNA bound to the dimer. By applying the QSSA to these two fast reactions, we obtain the sQSS solutions for $P_2$ and $D_A$ in terms of $P$: $P_2 (P)=P^2 /K_1$ and $ D_A (P)=\frac{K_2}{K_2 +P_2 (P)}$, where $K_1 = k_{b1}/k_{f1}$ and $K_2 =  k_{b2}/k_{f2}$. If we define $T \equiv P + 2P_2 + D_R$ and assume $D_R \ll T$, we obtain the tQSS solutions for $P_2$ and $D_A$ in terms of $T$: $P_2 (T)\approx \frac{K_1 + 4T - \sqrt{K^2_1 + 8K_1 T}}{2}$ and $D_A (T)=\frac{K_2}{K_2 + P_2 (T)}$. We use these QSS solutions to derive the propensity functions for the stochastic sQSSA and tQSSA. When the sensitivities of the sQSS solutions are much larger than those of the tQSS solutions (Fig.\ \ref{fig:dim}A), the stochastic sQSSA produces much larger errors in the average value of the fast variables, $P_2$ and $D_A$ than the stochastic tQSSA (Fig.\ \ref{fig:dim}C and E). As a result, the stochastic sQSSA results in a larger error in the CV of the slow variable, $n_M$, than the stochastic tQSSA (Fig.\ \ref{fig:dim} G). When the sensitivities of the sQSS solutions are reduced by changing parameters (Fig.\ \ref{fig:dim}B), the stochastic sQSSA more accurately predicts the average value of fast variables, $P_2$ and $D_A$ (Fig.\ \ref{fig:dim}D and F). Hence, the relative error in the CV of the slow variable, $n_M$, decreases (Fig.\ \ref{fig:dim}H).

\begin{figure} 
\centerline{\includegraphics[width=3in]{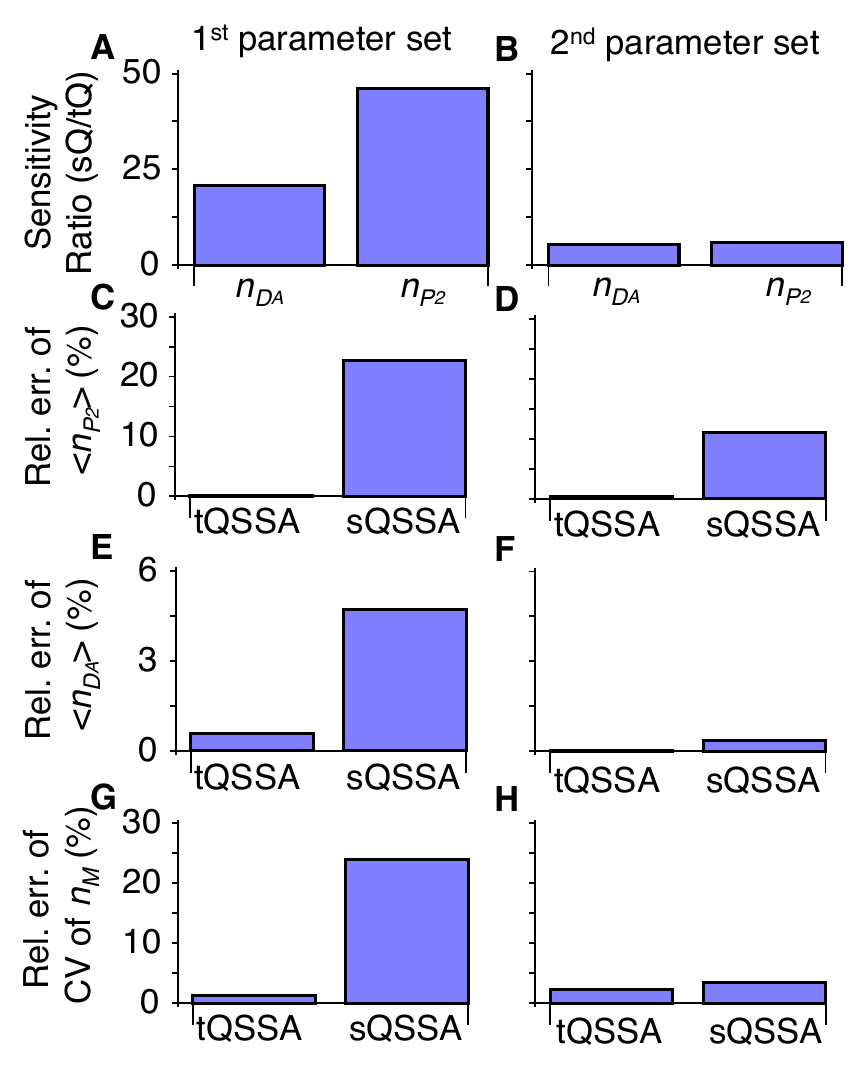}}
\caption{\label{fig:dim} Genetic negative feedback loop with protein dimerization.  (A, B) $\frac{dn_{D_A} /dn_P}{dn_{D_{A}}/dn_T}$ and $\frac{dn_{P_2} /dn_P}{dn_{P_{2}}/dn_T}$, ratios between the sensitivities of sQSS and tQSS solutions for $n_{D_A}$ and $n_{P_2}$ at equilibrium of two parameter sets (see Table S7 for parameters).  (C-F) Relative errors of the averages of fast species simulated using the stochastic QSSA. Here, $\left< n_D \right>$ and $\left< n_{P_2} \right>$ are the average of $n_D$ and $n_F$ at equilibrium. (G, H) Relative errors of CV of a slow specie, $n_M,$ at equilibrium. The results were obtained from  50,000 independent simulations. Each simulation is run until 5000s for the first parameter set and 15000s for the second parameter set of reaction time to ensure the system in the stationary state.}
\end{figure}

\section*{Linear noise approximation under slow timescales}
We have shown that the accuracy of the stochastic QSSA depends on both the sensitivity of the QSS solution and the variance of fast species (Eqs.\ \ref{eq:tes}-\ref{eq:sest}). However, the variance of fast species is usually unknown. Here, we derive the error in the variance of slow species simulated with the stochastic QSSA without using the variance of fast species. For this, we use a LNA that allows the estimation of the variance of variables in a mono-stable system when the number of molecules is not too small \cite{Thomas2011, Thomas2012, Agarwal2012, Elf2003}. Thus, with LNA, we can estimate the variance of slow species in the stochastic QSSA and compare with the original system.   

Consider a two dimensional deterministic system that consists of a slow species, $S,$ and a fast species, $F$,
\begin{equation}
\begin{array}{l}
\displaystyle \dot{S}=u(S,F),\\
\displaystyle \dot {F}=v(S,F).
\end{array} 
\label{eq:lfu}
\end{equation}
If the system is monostable, the corresponding LNA is given by
\begin{eqnarray}
\dot{\eta_{S}}&=&u_{S}\eta_{S}+u_{F}\eta_{F}+\frac{1}{\sqrt \Omega}S_{S}\sqrt{A} \vec{\Gamma}(t)\\
\dot{\eta_{F}}&=&v_{F}\eta_{S}+v_{F}\eta_{F}+\frac{1}{\sqrt \Omega}S_{F}\sqrt{A} \vec{\Gamma}(t),\ 
\end{eqnarray}
whose solutions, $\eta_{S}$ and $\eta_{F}$ provide approximations for the size of fluctuation of $S$ and $F$ from their steady state. $S_S$ and $S_F$ are stoichiometry matrices involving the variable $S$ and $F$, respectively; $A$ is a diagonal matrix whose entries are the elements of macroscopic rate functions; and $u_{S}, u_{F}, v_{S}$, and $v_{F}$ are components of the Jacobian at the steady state. Furthermore, $\vec{\Gamma}(t)$ is a vector of Gaussian noise whose elements, $\Gamma_i(t)$ for $i\in\left\{S,F\right\}$, satisfying
$
\left<\Gamma_i(t)\right>=0,$ and
$
\left<\Gamma_i(t)\Gamma_j(t')\right>=\delta_{ij}\delta(t-t'),
$
where $\delta_{ij}$ and $\delta(t)$ are the Kronecker and Dirac $\delta$-functions, respectively. Because the solutions of the LNA ($\eta_{S}$ and $\eta_{F}$) are multivariate Gaussian probability distributions, we can approximate the variance of $S$ and $F$, which is difficult to obtain from the original full CME  \cite{Thomas2011, Thomas2012, Agarwal2012, Elf2003}. Recently, Thomas \textit{et al.}\ \cite{Thomas2012} showed that when timescale separation holds, the effective stochastic description of intrinsic noise in the slow species can be described by the slow-scale LNA (ssLNA):
\begin{equation}\label{eq:lss}
\dot{\eta_{S}}=(u_{S}-u_{F}v^{-1}_{F}v_{S})\eta_{S}+\frac{1}{\sqrt \Omega}(S_{S}-u_{F}v^{-1}_{F}S_{F})\sqrt{A} \vec{\Gamma}(t).
\end{equation}
From this ssLNA, the variance of slow species ($\sigma_{S}$) can be derived by solving the Lyapunov equation \cite{Elf2003}, 
\begin{equation}\label{eq:lsv}
\sigma_{S}=\frac{S'AS'^{-1}}{2J\Omega},
\end{equation}
where $S'=S_{S}-u_{F}v^{-1}_{F}S_{F}$ and $J=u_{S}-u_{F}v^{-1}_{F}v_{S}$. 

Thomas \textit{et al.}\ \cite{Thomas2012} also derived the LNA corresponding to the reduced deterministic system with the QSSA:
\begin{equation}\label{eq:lqs}
\dot{\eta_{S}}=(u_{S}-u_{F}v^{-1}_{F}v_{F})\eta_{S}+\frac{1}{\sqrt \Omega}S_{S}\sqrt{A} \vec{\Gamma}(t).
\end{equation}
This is the LNA of the reduced stochastic model obtained via the stochastic QSSA. The diffusion term of this LNA does not have $u_{F}v^{-1}_{F}S_{F}$ in contrast to the ssLNA. Thus, the LNA corresponding to the stochastic QSSA (Eq. \ref{eq:lqs}) predicts the variance of slow species ($\sigma_{S}$), which is different from Eq. \ref{eq:lsv} of the ssLNA: 
\begin{equation}\label{eq:lqv}
\sigma_{S}=\frac{S_{S}AS_{S}^{-1}}{2J\Omega},
\end{equation}
where $J=u_{S}-u_{F}v^{-1}_{F}v_{S}$. Because $u_{F}v^{-1}_{F}S_{F}$ represents the contribution of the fast species to the variation of the slow species \cite{Thomas2012}, the difference between Eq. \ref{eq:lsv} and Eq. \ref{eq:lqv} indicates that the stochastic QSSA does not include the contribution of fast species to the variation of slow species. This is consistent with our moment analysis, which shows that the error of the stochastic QSSA stems from ignoring the variance of the fast species. Furthermore, $u_{F}v^{-1}_{F}$, which determines the error in $\sigma_{S}$ (Eq. \ref{eq:lqv}) simulated with the stochastic QSSA, can be directly calculated from the Jacobian of the deterministic system. Thus, by calculating $u_{F}v^{-1}_{F}$ of the deterministic system used in the tQSSA or sQSSA, we can estimate the accuracy of $\sigma_{S}$ simulated with the stochastic tQSSA or sQSSA. 

If we define $T \equiv S+F$, then from Eq.\ \ref{eq:lfu}, it follows that
\begin{eqnarray}
\dot{T}&=&u(S,F)+v(S,F) \equiv \overline u(T,F)\\
\dot{F}&=&v(S,F) \equiv \overline v(T,F).\
\end{eqnarray}
Then, the error in the diffusion term of the LNA corresponding to the stochastic tQSSA will be $\overline u_{F}\overline v^{-1}_{F}$. Implicit differentiation of the tQSS equation, $\overline v(T,F)=0$, gives $\frac{dF(T)}{dT}=-\frac{\overline v_T}{\overline v_{F}}$. From this, we can find the error in the diffusion term of the LNA corresponding to the stochastic tQSSA.  
\begin {equation}\label{eq:lte}
\frac{\overline u_F}{\overline v_{F}}=-\frac{\overline u_F}{\overline v_T}\frac{dF(T)}{dT}.
\end {equation}
Note that the error depends on the sensitivity of the tQSS solution, $\frac{dF(T)}{dT}$. In the example of Michaelis-Menten enzyme kinetics (Eqs.\ \ref{eq:ezfs}-\ref{eq:ezf}), the right side of Eq.\ \ref{eq:lte} becomes $\frac{k_p}{k_f E}\frac{dC(T)}{dT}$. This will be small because $\frac{dC(T)}{dT} \leq 1$ and $\frac{k_p}{k_f E} \ll 1$ due to timescale separation. This indicates that the stochastic tQSSA will accurately approximate the variance of slow species as long as timescale separation holds (Fig.\ \ref{fig:enz}).

In a similar way, we can derive the error of diffusion term in the LNA corresponding to the stochastic sQSSA (see supplementary information for details):
\begin {equation}\label{eq:lse}
-\frac{u_F+v_F}{v_{S}} \frac{dF(S)}{dS}.
\end {equation}
The error in the diffusion term of the stochastic sQSSA also depends on the the sensitivity of the sQSS solution,$ \frac{dF(S)}{dS}$. In the example of Eqs. \ref{eq:ezfs}-\ref{eq:ezf}, Eq.\ \ref{eq:lse} becomes $\frac{k_p+k_s}{k_f E}\frac{dC(S)}{dS}$. Due to timescale separation,$\frac{k_p+k_s}{k_f E_T} \ll1$. However, in contrast to $\frac{dC(T)}{dT} \le 1$, $\frac{dC(S)}{dS}$ can be very large depending on the parameter choice. Thus, even with timescale separation, the stochastic sQSSA cannot provide an accurate approximation for the variance of slow species if $\frac{dC(S)}{dS}$ is large. Furthermore, from Eqs.\ \ref{eq:lte} and \ref{eq:lse}, we can show that the ratio between these errors depends on $\frac{dF(S)/dS}{dF(T)/dT}$ similar to Eq.\ \ref{eq:mra} (see supplementary information for details). In summary, LNA analysis shows that the sensitivity of the QSS solution and timescale separation determine the error in the variance of slow species simulated with the stochastic QSSA.
\vspace*{-3pt}

\section*{Discussion}
Various deterministic QSSAs have been used to reduce ODE models of biochemical networks \cite{Menten1913,Briggs1925, Segel1989, Kepler2001, Schnell2000, Bennett2007, Tzafriri2003, Ciliberto2007, Kumar2011}. Recently, the macroscopic reaction rates obtained using deterministic QSSAs have  been used to derive approximate propensity functions for discrete stochastic simulations of slowly changing species (Fig.\ \ref{fig:sum}). Since this stochastic QSSA does not simulate rapidly fluctuating species, it greatly increases computation speed. The implicit assumption underlying this approach is that the stochastic QSSA is valid whenever its deterministic counterpart is valid, {\it i.e.}\ whenever timescale separation holds \cite{Gonze2002, Rao2003, Kim2012, Gonze2011a, Sanft2011}. If this were true, both the stochastic pQSSA and tQSSA would be equally accurate since their deterministic counterparts are dynamically equivalent (Fig.\ \ref{fig:det}). However, our simulations show that this is not always true and the stochastic tQSSA is more accurate than the stochastic pQSSA (Fig.\ \ref{fig:sto}A). 

We find that the accuracy of the stochastic QSSA is determined not only by timescale separation, but also the sensitivity of the QSS solution, which relates the fast species and the slow species at quasi-equilibrium (Fig.\ \ref{fig:sto}B). Specifically, our analysis of the moment equations shows that the sensitivity of QSS solutions determines how accurately the propensity functions obtained with the stochastic QSSA approximate the effective propensity functions obtained via the ssSSA (Fig.\ \ref{fig:enz}). This indicates that the propensity functions obtained from non-elementary reaction rate functions ({\it e.g.}\ Hill function) are accurate only when their sensitivity is low, which provides a novel condition for the validity of the stochastic QSSA. The error in the stochastic QSSA also depends on the variance of fast species, which is usually unknown. Therefore, low sensitivity does not guarantee the accuracy of the stochastic QSSA if the variance of fast species is too large. To address this problem, we also derived the error for the stochastic QSSA using LNA and noted that it does not depend on the variance of fast species. We showed that for a  mono-stable two dimensional system, the low sensitivity of the QSS solution is a  sufficient condition for the accuracy of stochastic QSSA as long as timescale separation holds. It will be interesting to test whether the low sensitivity of the QSS solution is a sufficient condition in more complex systems.

Whereas the stochastic QSSA uses the QSS solutions to approximate the average values of fast species \cite{Gillespie2007, Rao2003, Barik2008, Macnamara2008, Cao2005}, other methods (e.g.\ recursion relations) have been proposed to estimate the averages of fast species \cite{Goutsias2005, Cao2005, Agarwal2012}. These other methods could be used as alternatives when the stochastic QSSA is inaccurate (i.e.\ if the sensitivity of QSS solution is large). Finally, while the stochastic tQSSA is more accurate than the sQSSA or pQSSA, it is often difficult to find a closed form of the tQSS solution, and it needs to be calculated numerically \cite{Ciliberto2007, Kumar2011}. It will be important to understand how numerical calculation of the tQSS solutions affects computation time when using the stochastic tQSSA.


\section*{Acknowledgments}
We thank Hye-won Kang for valuable discussions and comments for this work. This work was funded by the NIH, through the joint NSF/NIGMS Mathematical Biology Program grant R01GM104974 (MRB and KJ), NSF grant DMS-1122094 (KJ), the Robert A.\ Welch Foundation grant C-1729 (MRB), and NSF grant DMS-0931642 to the Mathematical Biosciences Institute (JKK). 

\bibliographystyle{biophysj}


\end{document}


\title{The validity of quasi steady-state approximations in discrete stochastic simulations}

\author{Jae Kyoung Kim,$^1$ Kre\v simir Josi\' c,$^{2,3,\ast}$ and Matthew R.\ Bennett$^{4,5,\ast}$}

\maketitle

\noindent
{\footnotesize
{1}: Mathematical Biosciences Institute, The Ohio State University, Columbus, OH, 43210;\\
{2}: Department of Mathematics, University of Houston, Houston, TX, 77204;\\
{3}: Department of Biology and Biochemistry, University of Houston, Houston, TX, 77204;\\
{4}: Department of Biochemistry \& Cell Biology, Rice University, Houston, TX, 77005;\\
{5}: Institute of Biosciences and Bioengineering, Rice University, Houston, TX, 77005;\\
{*}: Correspondence: josic@math.uh.edu or matthew.bennett@rice.eduy \\}

\begin{abstract}
In biochemical networks, reactions often occur on disparate timescales and can be characterized as either ``fast'' or ``slow.'' The {\it quasi-steady state approximation} (QSSA) utilizes  timescale separation to project models of biochemical networks onto lower-dimensional slow manifolds. As a result, fast elementary reactions are not modeled explicitly, and their effect is captured by non-elementary reaction rate functions ({\it e.g.}\ Hill functions). The accuracy of the  QSSA applied to deterministic systems  depends on how well timescales are separated. Recently,  it has been proposed to use the non-elementary rate functions obtained via the deterministic QSSA to define propensity functions in stochastic simulations of biochemical networks. In this approach, termed the {\it stochastic QSSA}, fast reactions that are part of non-elementary reactions are not simulated, greatly reducing computation time. However, it is unclear when the stochastic QSSA provides an accurate approximation of the original stochastic simulation. We show that, unlike the deterministic QSSA, the validity of the stochastic QSSA does not follow from timescale separation alone, but also depends on the sensitivity of the non-elementary reaction rate functions to changes in the slow species. The stochastic QSSA becomes more accurate when this sensitivity is small. Different types of QSSAs result in non-elementary functions with different sensitivities, and the total QSSA 
results in less sensitive functions than the standard or the pre-factor QSSA. We prove that, as a result, the stochastic QSSA becomes more accurate when non-elementary reaction functions are obtained using the total QSSA. Our work provides a novel condition for the validity of the QSSA in stochastic simulations of biochemical reaction networks with disparate timescales. 
\end{abstract}

\markboth{Kim et al.}{QSSA in stochastic simulations}

\textbf{Note: This pre-print has been accepted for publication in Biophysical Journal. The final copyedited version of this paper will be available at www.biophyj.org }
\section*{The error of QSS solutions in approximating the average value of fast species}
In our model (Fig. 2), the reversible binding between free repressor protein and DNA,
\begin{equation}\label{eq:bu}
F+D_{A} {\longleftrightarrow} D_R,
\end{equation}
is much faster than other reactions. Therefore, the propensity function governed by this fast reaction, $\alpha_{M}n_{D_{A}}$ can be approximated by an effective propensity function, $\alpha_{M}\left< n_{D_{A}}\right>$~\cite{Gillespie2007, Cao2005, Cai2007} under timescale separation (see Eq. 28 for the detail definition of $\left< n_{D_{A}}\right>$) ~\cite{Gillespie2007, Cao2005, Cai2007}. In this section, we estimate the difference between $\left<n_{D_{A}}\right>$ and QSS solutions (Eqs. 25 and 26) using a moment relation~\cite{Sanft2011}. The steady-state CME for the reactions of Eq.\ \ref{eq:bu} in isolation yields 
\begin{equation}\label{eq:mfu}
\left<k_{f} n_F n_{D_{A}}/\Omega\right>=\left< k_{b}n_{D_R}\right>.
\end{equation}
Using the definitions $n_R=n_F +n_{D_R}$ and $n_{D_T} =n_{D_R} +n_{D_A}$, Eq.\ \ref{eq:mfu} becomes $\left<(n_R-n_{D_R})(n_{D_{T}}-n_{D_R})/\Omega\right>=\left< \bar{K_{d}}n_{D_R}\right>$, where $\bar{K_{d}}=k_{b}/k_{f}$. Note that $n_{D_{T}}$ is constant and the slow variable, $n_{R}$, is in quasi-equilibrium. Therefore $n_{R}$ can be treated as a constant and determines the state $\mathbb{S}$ in Eq. 28, while $n_{D_T}$ and $n_{R}$ can be factored out of the average to obtain $\left<{n_{D_R}}^2\right>-(n_{D_{T}}+n_R+\bar{K_{d}}\Omega)\left<n_{D_R}\right>+n_{D_{T}}n_R=0$. Using the relation, $\left<n_{D_R}^2\right>=\left<n_{D_R}\right> ^2+Var(n_{D_R})$, we get
\begin{eqnarray}
\left<n_{D_R}\right>^2-(n_{D_{T}}+n_{R}+\bar{K_{d}}\Omega)\left<n_{D_R}\right>+n_{D_{T}}n_{R}+Var(n_{D_R})=0.\label{eq:mtf}\
\end{eqnarray}
We next close this equation at first moment by truncating $Var(n_{D_R})$. We can relate the solution of this truncated equation  to the tQSS solution obtained using the deterministic tQSSA (Eq. 16), if
we  rewrite in terms of $n_{D_R}$, rather than  $\left<n_{D_R}\right>$ and replace $\bar{K_d}$ by $K_d$ because $k_f \gg k_p$ and $\bar{K_d} \approx K_d$:
\begin{equation}\label{eq:mtq}
tQ(n_{D_R},n_{R}) \equiv n_{D_R}^2 -(n_{D_{T}}+n_{R}+K_{d}\Omega)n_{D_R}+n_{D_{T}}n_{R}=0.
\end{equation}
Dividing by $\Omega^2$ gives Eq. 16. We used the solution of Eq. \ref{eq:mtq} to define the propensity function for transcription in the stochastic tQSSA (Eq. 26). Thus, the relation obtained using the tQSSA (Eq. \ref{eq:mtq}) agrees with the moment equation for $\left<n_{D_R}\right>$ {\it under the moment closure assumption}, that is, under the assumption that the first moment does not depend on higher moments. This indicates that the error in the stochastic QSSA is mainly due to ignoring the variance of fast variables, which will vanish along with random fluctuations in the limit of large system size. Furthermore, $Var(n_{D_R})$ is typically of order $\left<n_{D_R}\right>$, so $Var(n_{D_R}) \ll n_{D_{T}}n_{R}$ \cite{Sanft2011}. Thus, $n_{D_R}(n_R)$, the solution of Eq.\ \ref{eq:mtq}  should be close to $\left<n_{D_R}\right>$, the solution of Eq.\ \ref{eq:mtf}. This  explains the accuracy of the stochastic tQSSA (Fig.\ 3). 

To make this argument concrete, we estimate the difference between $\left<n_{D_R}\right> $ and $n_{D_R}(n_R)$. Because $Var(n_{D_R}) \ll n_{D_{T}}n_{R}$ and $\bar{K_d} \approx K_d$, we can approximate the solution of Eq.\ \ref{eq:mtf} using a Taylor expansion,
\begin{equation}\label{eq:mte}
\left<n_{D_R}\right> \approx n_{D_R}(n_R)-\frac{Var(n_{D_R})}{2n_{D_R}(n_R)-(n_{D_{T}}+n_R+K_{d}\Omega)},
\end{equation}
good to first order in $Var(n_{D_R})$. By implicit differentiation of Eq.\ \ref{eq:mtq}, we  obtain
\begin{eqnarray}
\frac{dn_{D_R}(n_R)}{dn_R}&=&-\frac{\partial tQ(n_{D_R},n_R)/\partial n_R} {\partial tQ(n_{D_R},n_R)/\partial n_{D_R}}\nonumber\\
	&=&-\frac{n_{D_{A}}(n_{R})}{2n_{D_R}(n_R)-(n_{D_{T}}+n_R+K_{d}\Omega)}.\
\end{eqnarray}
Substituting this equation into Eq. \ref{eq:mte} gives
\begin{equation}
\left<n_{D_R}\right> \approx n_{D_R}(n_R)+\frac{Var(n_{D_R})}{n_{D_{A}}(n_R)} \frac{dn_{D_R}(n_R)}{dn_R}.
\end{equation}
Since $n_{D_{A}}=n_{D_{T}}-n_{D_R}$, $Var(n_{D_R})=Var(n_{D_{A}})$ and $dn_{D_R}/dn_R=-dn_{D_{A}}/dn_R$, so 
\begin{equation}\label{eq:mtet}
\left<n_{D_{A}}\right> \approx n_{D_{A}}(n_R)+\frac{Var(n_{D_{A}})}{n_{D_{A}}(n_R)} \frac{dn_{D_{A}}(n_R)}{dn_R}.
\end{equation}
The error depends on the Fano factor of the fast species, $\frac{Var(n_{D_{A}})}{n_{D_{A}}(n_R)}$. This reflects the fact that the error is due to the truncation of variance of the fast species in Eq.\ \ref{eq:mtq}. The error magnitude depends on $\frac{dn_{D_{A}}}{dn_{R}}$, the sensitivity of the tQSS solution, which is small (Eq. 26). This explains the accuracy of the stochastic tQSSA (Fig. 3).

For the sQSSA and the pQSSA, however, $n_{D_A}$ is a function of $n_F$, but not $n_R$. We therefore estimate $\left<n_{D_{A}}\right>$ as a function of $n_F$ rather than $n_R$. 
Using the definition $n_{D_T} =n_{D_R} +n_{D_A}$, Eq.\ \ref{eq:mfu} becomes
\begin{equation}\label{eq:mfut}
\left< n_F n_{D_R}\right>+\bar{K_{d}}\Omega \left<n_{D_R}\right>-n_{D_{T}}\left<n_F\right>=0.
\end{equation}
Importantly, whereas $n_R$ changes slowly on fast timescales, $n_F$ depends on both fast and slow reactions, and can change rapidly. Thus, unlike $n_R$, $n_F$ cannot be treated as constant. We expand $\left< n_{F}n_{D_R}\right>$ using the property
\begin{eqnarray}
\left< n_{F}n_{D_R}\right>-\left< n_F\right>\left<n_{D_R}\right>& \equiv &Cov(n_F,n_{D_R})\nonumber\\
	&=&Cov(n_R-n_{D_R},n_{D_R})\nonumber\\
	&=&-Var(n_{D_R}),\label{eq:scov}\
\end{eqnarray}
The last equality follows from $Cov(n_R,n_{D_R})=0$ since $n_R$ is in quasi-equilibrium and is treated as a constant. Substituting Eq.\ \ref{eq:scov} into Eq.\ \ref{eq:mfut}, we obtain
\begin{equation} \label{eq:msf}
\left< n_F\right>\left<n_{D_R}\right>+\bar{K_{d}}\Omega \left<n_{D_R}\right>-n_{D_{T}}\left< n_F\right>-Var(n_{D_R}) = 0.
\end{equation}
Again Eq.\ \ref{eq:msf} is related to the equation used to define $n_{D_R}$ as a function of $n_F$ in the stochastic sQSSA/pQSSA.  To see this, we close the equation at the first moment by truncating $Var(n_{D_R})$, and we replace $\bar{K_d}$ by $K_d$.  Writing this truncated equation in terms of $n_{D_R}$, rather than  $\left<n_{D_R}\right>$, we obtain:
\begin{equation}\label{eq:msq}
sQ( n_{D_R}, \left< n_F\right>) \equiv \left< n_F\right> n_{D_R} +K_{d}\Omega  n_{D_R} -n_{D_{T}}\left< n_F\right>=0,
\end{equation}
which has the solution $n_{D_R}(\left<n_F\right>)=\frac{n_{D_{T}}\left< n_F\right>}{\left< n_F\right>+ K_{d}\Omega}$. This solution agrees with the expression for $n_{D_R}(n_F)$ derived from the sQSS solution (Eq. 6) because $\left< n_F\right>$ approximates $n_F$ under slow timescale. Since $\bar{K_d} \approx K_d$,  it follows from Eqs.\ \ref{eq:msf}~and~\ref{eq:msq} that
\begin{equation} \label{eq:mse}
\left<n_{D_R}\right> \approx n_{D_R}(\left<n_F\right>)+\frac{Var(n_{D_R })}{\left<n_F\right>+ K_{d}\Omega}.
\end{equation}
Implicit differentiation of Eq.\ \ref{eq:msq} yields $\frac{dn_{D_R}(\left<n_F\right>)}{d\left<n_F\right>} = \frac{n_{D_{A}}(\left<n_F\right>)}{\left<n_F\right>+ K_{d}\Omega}$. Using this equation to substitute the denominator in Eq.\ \ref{eq:mse}, we obtain
\begin{equation} \label{eq:moment2}
\left<n_{D_R}\right> \approx n_{D_R}(\left<n_F\right>)+\frac{Var(n_{D_R} )}{n_{D_{A}}(\left<n_F\right>)}\frac{dn_{D_R}(\left<n_F\right>)}{d\left<n_F\right>}.
\end{equation}
From this, similar to above, we can derive
\begin{equation}\label{eq:mset}
\left<n_{D_{A}}\right> \approx n_{D_{A}}(\left<n_F\right>)+\frac{Var(n_{D_{A}})}{n_{D_{A}}(\left<n_F\right>)}\frac{dn_{D_{A}}(\left<n_F\right>)}{d\left<n_F\right>}.
\end{equation}
This estimates the error when the $n_{D_{A}}(\left<n_F\right>)$ derived from the sQSS solution (Eq. 7) is used to approximate $\left<n_{D_{A}}\right>$. Similar to the tQSS solution, the error again depends on $\frac{Var(n_{D_{A}})}{n_{D_{A}}(n_R)}$, the Fano factor of the fast species. This error is amplified by the sensitivity of the sQSS solution ($\frac{dn_{D_{A}}(\left<n_F\right>)}{d\left<n_F\right>}$), which can be large (Eq. 25). These results are generalized to any system in which reversible binding reactions are faster than other reactions.\\

\textbf{Theorem}. Assume that a  biochemical reaction network includes a reversible binding reaction with a dissociation constant $K_d=k_b/k_f$, 
\begin{equation}\label{eq:bur}
S+F {\longleftrightarrow} C.
\end{equation}

that is faster than the other reactions in the system.
Let $T\equiv S+C$ and $U\equiv F+C$. If $Var(n_C)\ll n_T n_U$, then $\left<n_C\right>$ and $\left<n_F\right>$ satisfy: 
\begin{eqnarray}
\left<n_C\right> &\approx& n_C(n_T)+\frac{Var(n_C)}{n_F(n_T)} \frac{dn_C(n_T)}{dn_T}\label{eq:tes}\\  
\left<n_F\right> &\approx& n_F(n_T)+\frac{Var(n_F)}{n_F(n_T)} \frac{dn_F(n_T)}{dn_T},\
\end{eqnarray}
where $n_C(n_T)$ is the solution of the tQSS equation, $n_C^2-(n_U+n_T+K_d \Omega)n_C+n_Un_T=0$, and  $n_F(n_T)=n_U-n_C(n_T)$. 
Similarly,
\begin{eqnarray}
\left<n_C\right> &\approx& n_C(\left<n_S\right>)+\frac{Var(n_C)}{n_F(\left<n_S\right>)}\frac{dn_C(\left<n_S\right>)}{d\left<n_S\right>}\label{eq:ses}\\
\left<n_F\right> &\approx& n_F(\left<n_S\right>)+\frac{Var(n_F)}{n_F(\left<n_S\right>)}\frac{dn_F(\left<n_S\right>)}{d\left<n_S\right>},\label{eq:sest}\
\end{eqnarray}
where $n_C(\left<n_S\right>)$  is the solution of the sQSS equation, $(\left<n_S\right>+K_d\Omega)n_F + n_U\left<n_S\right>=0$, and $n_F(\left<n_S\right>)=n_U-n_C(\left<n_S\right>)$. 
\begin{proof}
The proof follows from the derivations above, because the reversible binding (Eq. \ref{eq:bur}) is faster than other reactions, assuming quasi-equilibrium leads to the relation between moments given in Eq.\ \ref{eq:mfu}. Furthermore, due to the cancelation of binding and unbinding reactions, $T$ and $U$ evolve on the slow timescale, and can be considered constant on fast timescales.  We can therefore use moment expansions equivalent to Eq.\ \ref{eq:mtf} and Eq.\ \ref{eq:msf}. Finally, we can follow the derivation of Eq.\ \ref{eq:mtet} and Eq.\ \ref{eq:moment2} by using a Taylor expansions to estimate the error in the QSS equations resulting from the truncation of higher moments in Eq.\ \ref{eq:mtf} and Eq.\ \ref{eq:msf}.\\
\end{proof}

\section*{Linear noise approximation corresponding to the stochastic sQSSA}
Let us estimate the error in the diffusion term of the LNA corresponding to the stochastic sQSSA. When the system is reduced with the sQSSA, the fast reactions in $\dot {S}$ (Eq. 50) are switched to the slow reactions by using the QSS equation, $v(S,F)=0$. For instance, $-k_f SE + k_b C$ in Eq. 37 is switched to $-k_p C(S)$ in Eq. 39 by using the QSS equation, $k_f SE - k_b C - k_p C=0$. This means that the LNA corresponding to the reduced system with the sQSSA will have $S_{S}+S_{F}$ rather than $S_{S}$ in the diffusion term of Eq. 55. Due to this change, the difference in diffusion terms of Eq. 54 and Eq. 56 becomes $(\frac{u_F}{v_{F}}+1)S_F$ rather than $\frac{u_F}{v_{F}}S_F$. Thus, the error in the diffusion term becomes
\begin {equation}\label{eq:lses}
\frac{u_F}{v_{F}}+1=-\frac{u_F+v_F}{v_{S}} \frac{dF(S)}{dS},
\end {equation}
where the equality comes from $\frac{dF(S)}{dS}=-\frac{v_F}{v_S}$, which is obtained by applying the implicit function theorem to the QSS equation $v(S,F)=0$. To compare the errors of diffusion term in the tQSSA and the sQSSA, let us change the variable $S$ to $T$ in Eq.\ \ref{eq:lses} by using following relationships: 
\begin{eqnarray}
u(S,F)&=&\overline u(T,F) - \overline v(T,F)\nonumber\\
	&\Longrightarrow& u_F = \overline u_T +\overline u_F - \overline v_T - \overline v_F\\
v(S,F)&=&\overline v(T,F)\nonumber\\
	&\Longrightarrow& v_S =\overline v_T , v_F=  \overline v_T + \overline v_F.\
\end{eqnarray}
Then, Eq.\ \ref{eq:lses} becomes
\begin{equation}\label{eq:lsest}
-\frac{u_F+v_F}{v_{S}} \frac{dF(S)}{dS}=-\frac{\overline u_{T}+\overline u_{F}}{\overline v_{T}} \frac{dF(S)}{dS},
\end {equation}
and, from Eqs. 59 and \ref{eq:lsest}, the ratio between errors of diffusion term in the sQSSA and the tQSSA becomes
\begin {equation}
\frac{\overline u_{T}+\overline u_{F}}{\overline u_F} \frac{dF(S)/dS}{dF(T)/dT}.
\end {equation}

\bibliographystyle{biophysj}
\bibliography{QSS_supp_v3.bbl}

\pagebreak
\section*{Supplementary Tables}
\begin{table}[h!]
\begin{center}
\caption{Parameters of genetic negative feedback loop model}
  \begin{tabular}{| l | l | l |} 
  \hline
   Name & Description &Value \\ \hline
  $\alpha_M$ & Transcriptional rate constant for  $M$ & 15.1745/hr  \\
  $\alpha_P$ & Translational rate constant for  $P$ & 1/hr  \\
  $\alpha_F$ & Production rate constant for $F$ & 1/hr  \\
 $\beta_M$ & Degradation rate constant for  $M$ & 1/hr  \\
 $\beta_P$ & Degradation rate constant for  $P$ & 1/hr  \\
$\beta_F$ & Degradation rate constant for  $F$ & 1/hr  \\
 $k_f$ & Binding rate constant for $F$ to $D_A$ & 200/nM hr \\
 $k_b$ & Unbinding rate constant for $D_R$& 50/hr \\
 $D_T$ & The concentration of total DNA &164.75nM\\
 \hline  
\end{tabular}
\end{center}
\end{table}

\begin{table}[h!]
\begin{center}
\caption{Propensity functions of reactions in full system of genetic negative feedback loop}
  \label{tab:prof}
  \begin{tabular}{| c | c |} 
  \hline
Reaction & Propensity function\\ \hline
 $D_A  \xrightarrow{} D_A + M$ & $\alpha_M n_{D_A}$  \\
 $M \xrightarrow{} M+ P$ & $\alpha_P n_{M}$ \\
 $P \xrightarrow{} P+ F$ & $\alpha_F n_{P}$ \\
 $M  \xrightarrow{}  \phi $ & $\beta_M n_{M}$  \\
 $P  \xrightarrow{}  \phi $ & $\beta_P n_{P}$ \\
 $F  \xrightarrow{}  \phi $ & $\beta_F n_{F}$ \\ 
 $D_R  \xrightarrow{}  D_A $ & $\beta_F n_{D_R}$ \\ 
 $D_A + F \xrightarrow{} D_R$ & $k_f n_{D_A}n_F/ \Omega$ \\
 $D_R \xrightarrow{} D_A + F$ & $k_b n_{D_R}$\\ 
 \hline  
\end{tabular}
\end{center}
\end{table} 
\begin{table}[h!]
\begin{center}
\caption{Propensity functions of genetic negative feedback loop obtained using the stochastic sQSSA}
  \label{tab:pros}
  \begin{tabular}{| c | c |} 
  \hline
Reaction & Propensity function\\ \hline
 $\phi  \xrightarrow{}  M$ & $\frac{\alpha_M n_{D_T}K_d \Omega}{n_F + K_d \Omega } $  \\
 $M \xrightarrow{} M+ P$ & $\alpha_M n_{M}$ \\
 $P \xrightarrow{} P+ F$ & $\alpha_P n_{P}$ \\
 $M  \xrightarrow{}  \phi $ & $\beta_M n_{M}$  \\
 $P  \xrightarrow{}  \phi $ & $\beta_P n_{P}$ \\
 $F  \xrightarrow{}  \phi $ & $\beta_F (n_F +\frac{n_{D_T}n_{F}}{n_F + K_d \Omega})$ \\ 
  \hline  
\end{tabular}
\end{center}
\end{table} 
\begin{table}[h!]
\begin{center}
\caption{Propensity functions of genetic negative feedback loop obtained using the stochastic tQSSA}
  \label{tab:prot}
  \begin{tabular}{| c | c |} 
  \hline
Reaction & Propensity function\\ \hline
 $\phi  \xrightarrow{}  M$ & \tiny{ $\frac{\alpha_M}{2} (n_{D_T}-n_R - K_d \Omega  - \sqrt{(n_{D_T}-n_R - K_d \Omega)^2 +4n_{D_T}K_d \Omega})$ } \\
 $M \xrightarrow{} M+ P$ & $\alpha_M n_{M}$ \\
 $P \xrightarrow{} P+ R$ & $\alpha_P n_{P}$ \\
 $M  \xrightarrow{}  \phi $ & $\beta_M n_{M}$  \\
 $P  \xrightarrow{}  \phi $ & $\beta_P n_{P}$ \\
 $R  \xrightarrow{}  \phi $ & $\beta_F n_R$ \\ 
  \hline  
\end{tabular}
\end{center}
\end{table} 
\begin{table}[h!]
\begin{center}
\caption{Propensity functions of genetic negative feedback loop obtained using the stochastic pQSSA}
  \label{tab:pros}
  \begin{tabular}{| c | c |} 
  \hline
Reaction & Propensity function\\ \hline
 $\phi  \xrightarrow{}  M$ & $\frac{\alpha_M n_{D_T}K_d \Omega}{n_F + K_d \Omega } $  \\
 $M \xrightarrow{} M+ P$ & $\alpha_M n_{M}$ \\
 $P \xrightarrow{} P+ F$ & $\alpha_P n_{P} / (1+\frac{n_{D_T}K_d \Omega}{(n_F + K_d \Omega)^2})$ \\
 $M  \xrightarrow{}  \phi $ & $\beta_M n_{M}$  \\
 $P  \xrightarrow{}  \phi $ & $\beta_P n_{P}$ \\
 $F  \xrightarrow{}  \phi $ & $\beta_F (n_F +\frac{n_{D_T}n_{F}}{n_F + K_d \Omega})/ (1+\frac{n_{D_T}K_d \Omega}{(n_F + K_d \Omega)^2})$ \\ 
  \hline  
\end{tabular}
\end{center}
\end{table} 

\begin{table}[h!]
\begin{center}
\caption{Parameters of enzyme kinetics model}
  \begin{tabular}{| l | l | l |l |} 
  \hline
   Name & Parameter description &1st set & 2nd set\\ \hline
  $k_f$ & Binding rate constant for $E$ to $S$& 0.01/nM s & 0.017/nM s \\
  $k_b$ & Unbinding rate constant for $C$ & 0.02/s & 0.03/s\\
 $k_p$ & Production rate constant for $P$ & 0.002/s & 0.0016/s\\
 $k_s$ & Conversion rate constant for $P$ to $S$ & 0.001/s & 0.0007/s\\
 $E_T$ & The concentration of total enzyme & 25nM & 40nM\\
 $S_T$ & The concentration of total substrate & 100nM & 78nM\\
  \hline  
\end{tabular}
\end{center}
\end{table}

\begin{table}[h!]
\begin{center}
\caption{Parameters of genetic negative feedback loop with dimerization model}
  \begin{tabular}{| l | l |l|l|} 
  \hline
   Name & Parameter description &1st set & 2nd set\\ \hline
  $k_{f1}$ & Binding rate constant for $P$ to $P$ & 0.025/nM s & 0.25/nM s \\
  $k_{b1}$ & Unbinding rate constant for $P_2$ & 0.5/s & 0.1/s\\ 
  $k_{f2}$ & Binding rate constant for $P_2$ to $D_A$ &0.012/nM s & 0.025/nM s \\
  $k_{b2}$ & Unbinding rate constant for $D_R$ & 0.9/s & 0.9/s\\
 $\alpha_M$ & Transcription rate constant for  $M$ & 0.0078/s & 0.06175/s\\
 $\alpha_P$ & Translational rate constant for $P$ &0.043/s & 0.00043/s\\
 $\beta_M$ & Degradation rate constant for $M$ & 0.0039/s & 0.0039/s\\
 $\beta_P$ & Degradation rate constant for $P$ & 0.0007/s & 0.0007/s\\
 $D_T$ & The concentration of total DNA  &1nM & 1nM\\
  \hline  
\end{tabular}
\end{center}
\end{table}